\documentclass[preprint]{aastex}

\usepackage{graphics}

\newcommand{\kms}{km~s$^{-1}$}
\newcommand{\msol}{$M_{\sun}$}
\newcommand{\um}{$\mu$m}

\newcommand{\aall}{$\alpha_{2-24\mu{\rm m}}$}
\newcommand{\airac}{$\alpha_{\rm IRAC}$}
\newcommand{\ntwoone}{$N_{\rm ClassII}/N_{\rm ClassI}$}
\newcommand{\vlsr}{v_{\rm LSR}}
\newcommand{\nco}{$N({\rm ^{13}CO})$}
\newcommand{\tex}{$T_{\rm ex}$}
\newcommand{\tmb}{$T_{\rm mb}$}

\shorttitle{Star Formation in the IRDC G53.2}
\shortauthors{Kim, Koo, and Davis}

\begin{document}
\title{Star Formation Activity in the Long, Filamentary
Infrared Dark Cloud G53.2}

\author{Hyun-Jeong Kim\altaffilmark{1} and Bon-Chul Koo\altaffilmark{2}}
\affil{Department of Physics and Astronomy, 
Seoul National University,\\ 
Seoul 151-742, Korea}
\and
\author{Christopher J. Davis\altaffilmark{3}}
\affil{Astrophysics Research Institute, 
Liverpool John Moores University, \\
Liverpool, L3 5RF, UK}
\altaffiltext{1}{hjkim@astro.snu.ac.kr}
\altaffiltext{2}{koo@astro.snu.ac.kr}
\altaffiltext{3}{c.j.davis@ljmu.ac.uk}

\begin{abstract}
We present star formation activity in the infrared dark cloud
(IRDC) G53.2, a remarkable IRDC located at Galactic coordinates 
$(l,b)\sim(53^{\circ}.2,\,0^{\circ}.0)$ based on the census of young 
stellar object (YSO) candidates. IRDC G53.2 was previously identified 
as several IRDCs in mid-IR images, but it is in fact 
a long ($\gtrsim$45 pc) cloud, well consistent with a CO cloud 
at $v\sim23$~\kms\ (or at $d\sim$1.7 kpc). We present a point-source 
catalog of IRDC G53.2 that contains $\sim$370 sources from our 
photometry of the {\it Spitzer} MIPS 24~\um\ data and Galactic
Legacy Infrared Mid-Plane Survey Extraordinaire Catalog. 
The classification of the identified sources based on their spectral index and 
control field analysis to remove field star contamination reveals that IRDC G53.2 
is an active star-forming region with $\sim$300 YSO candidates. We compare 
the YSO classification based on spectral index, mid-IR colors, and 
the wavelength range used, which results in consistent classification, 
except for flat-spectrum objects, with some ambiguity between Class I and II. 
Comparison of the YSO population in IRDC G53.2 with those of other 
nearby star-forming clusters indicates that they are similar in age; 
on the other hand, stronger association with mid-IR stellar sources in 
IRDC G53.2 compared with other IRDCs indicates that IRDC G53.2 
is at a later evolutionary stage among IRDCs. Spatial distribution of 
the YSO candidates in IRDC G53.2 shows a good correlation 
with $^{13}$CO column density and far-IR emission, and earlier-class 
objects tend to be more clustered in the regions with higher density.
\end{abstract}


\section{Introduction}

Massive stars ($M \gtrsim 8$\msol) have a great influence in 
the interstellar medium and the galactic environment by providing 
ionizing photons and kinetic energy, enriching heavy elements, and so on.
Nonetheless, compared with low-mass stars, 
the formation process of massive stars still remains unclear
because direct observations of high-mass protostars are challenging 
owing to their rarity, long distances, and short lifetimes.
Infrared dark clouds (IRDCs), which are identified as silhouettes
against the bright Galactic background in mid-IR, are 
cold ($<$20~K) and very dense ($>$10$^5$~cm$^{-3}$) clouds 
with high column 
densities \citep[$\sim$10$^{21}$--10$^{23}$~cm$^{-2}$;][]
{carey98,carey00,egan98,pillai06,peretto10,ragan11}.
Thus, IRDCs are believed to be the precursors to massive stars and star clusters.
Millimeter/far-IR clumps/cores found in IRDCs with masses from a few tens to
thousands of solar masses support this \citep[e.g.,][]{rath06, henning10},
while the investigation of star formation in the IRDC at Galactic coordinates 
$(l, b) \sim (48^{\circ}.66,\, -0^{\circ}.30)$
using {\it Spitzer} data \citep{wiel08} showed a diverse distribution in
the mass of young stellar objects (YSOs), with no massive ($>$8~\msol) ones, 
although their sample size was small ($<$20).

Since IRDCs  were cataloged from 
the {\it Midcourse Space Experiment} ({\it MSX}) 8~\um\ 
data \citep[MSXDC catalog;][]{simon06a} 
and more recently from the {\it Spitzer}
Galactic Legacy Infrared Mid-Plane Survey Extraordinaire
(GLIMPSE; \citealt{benjamin03})%
\footnote{http://www.astro.wisc.edu/glimpse/glimpsedata.html}
8~\um\ data \citep{peretto09},
several studies on physical properties and star formation 
characteristics of IRDCs 
via multiwavebands from mid- and far-IR to radio 
have been published \citep[e.g.,][]
{rath06,simon06b,chambers09,battersby10,jackson10}.
Most studies have mainly focused on 
statistical properties of IRDCs/IRDC clumps themselves 
or star formation activities on clump or core scales
rather than on a stellar scale represented by mid-IR detection.
The IRDCs in the previous studies are composed of several clumps 
in a diverse mass range \citep{ragan09,henning10}, 
and as IRDC clumps evolve, they tend to be more associated 
with star formation tracers, including mid-IR (24~\um) stellar 
sources \citep{chambers09,battersby10}.
However, only a small fraction of IRDC clumps that have thus far been 
investigated are associated with 24~\um\ emission features \citep{ragan09}. 
The number of 24~\um\ stellar sources in IRDCs is usually very 
limited \citep{wiel08,henning10}, too few for a proper statistical analysis 
of star formation in individual IRDCs after the clumps (i.e., the precursors to 
massive stars and clusters) further fragment.
Therefore, an investigation on IRDCs with many young stars 
(i.e., large number of stellar sources detected in mid-IR) 
such as the one presented in this study 
and comparison with star formation activities in other 
well-studied star-forming regions are crucial 
to understanding the statistical properties of star formation
(e.g., YSO population, star formation rate, or 
initial mass function) in IRDCs. 

In this paper, we present star formation activity in an IRDC 
located at Galactic coordinates $(l, b) \sim (53^{\circ}.2,\, 0^{\circ}.0)$. 
This IRDC (IRDC G53.2, hereafter) is long and filamentary, extending 
$\gtrsim$1$^{\circ}$ in the mid-IR images as seen in Figure~\ref{fig1}, and 
it shows a number of bright mid-IR stellar sources distributed along 
its filamentary structure. 
IRDC G53.2 was previously identified as three separate IRDCs 
in the MSXDC catalog \citep[MSXDC G053.11+00.05, MSXDC G053.25+00.04,
MSXDC G053.31+00.00;][]{simon06a} as presented in Figure~\ref{fig1}, 
and as 56 IRDCs in the {\it Spitzer} IRDC catalog \citep{peretto09}; however,
we have found that these individual IRDCs in the previous two catalogs 
lying between $l \sim 53^{\circ}$ and 54$^{\circ}$ are 
well consistent with a CO cloud at $v \sim 23$ \kms.
Therefore, in this study, we consider these separate IRDCs 
in the mid-IR image to be associated, and we investigate 
the overall star formation activity in IRDC G53.2 using a 
catalog of YSO candidates constructed from {\it Spitzer} data.
This paper is organized as follows. In Section~2, 
we define IRDC G53.2 based on the associated molecular
cloud and present its physical parameters. 
In Section~3, we describe the point-source catalog (PSC) of 
IRDC G53.2, followed by selection and classification of 
YSO candidates in Section~4 and accounting for contaminations 
in Section~5.
We then discuss different schemes of YSO classification, 
the characteristics of star formation in IRDC G53.2 
in terms of the stellar population, and spatial distribution of 
the YSO candidates in Section~6.
In Section~7, we give our summary and conclusions.

\section{Molecular Cloud Associated with IRDC G53.2}

\begin{figure}
\begin{center}
\includegraphics[width=0.6\textwidth]{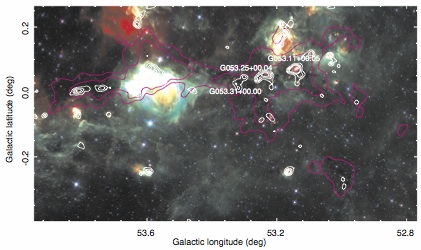}
\caption{Three-color image of IRDC G53.2 produced from 
{\it Spitzer} IRAC 5.8 $\micron$ (B), IRAC 8.0 $\micron$ (G), and 
MIPS 24 $\micron$ (R) images. 
The magenta contours are from the GRS $^{13}$CO $J=$~1--0 
integrated intensity map at $v=$ 15--30 \kms\ (see Section~2 
and Figure~\ref{fig2}).
The outermost contour level, which defines the boundary of 
IRDC G53.2, is $\int T_{mb} dv= 6.3$ K~\kms\ or 
the mean antenna temperature 
$\bar T_A(=0.48 \bar T_{mb})= 0.2$ K.
The Bolocam Galactic Plane Survey \citep{bolocam} 1.2 mm 
contours are also overlaid in white. The previously identified 
three IRDCs in the MSXDC catalog \citep{simon06a} are marked. 
\label{fig1}}
\end{center}
\end{figure}

\begin{figure}
\begin{center}$
\begin{array}{cc}
\includegraphics[width=0.4\textwidth]{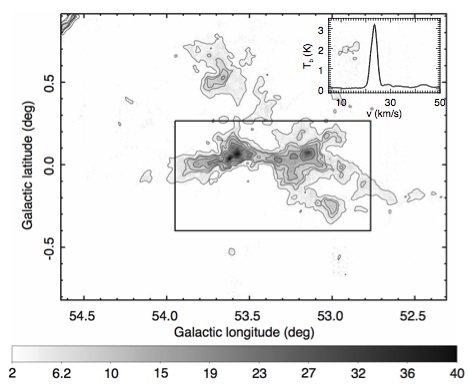} &
\includegraphics[width=0.4\textwidth]{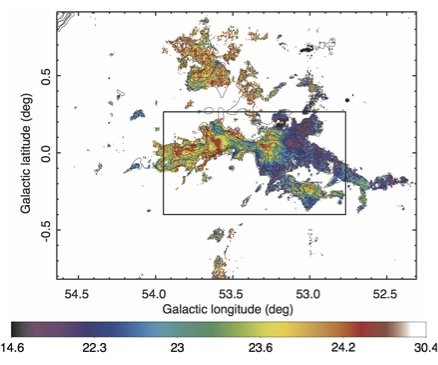}
\end{array}$ 
\end{center}
\caption{Left: $^{13}$CO $J=$~1--0 integrated intensity map of 
the IRDC G53.2 region. 
The velocity range is from $\vlsr=+15$ to +30~\kms.
The scale bar shows the integrated intensity scale, and 
the contour levels are drawn at 
$\int T_{mb} dv=$3.1, 6.3, 7.9, 11.0, and 15.7 K~\kms\
(or the mean antenna temperature 
$\bar T_A(=0.48 \bar T_{mb})=$0.1, 0.2, 0.25, 0.35, and 0.5 K).
The inset shows the average spectrum of the cloud, and
the square box marks the area of Figure~\ref{fig1}. 
Right: Mean velocity of the $^{13}$CO gas in IRDC G53.2. 
The velocity scale (\kms) is given by the scale bar at the bottom. 
The contour levels of the integrated intensity are overlaid.
\label{fig2}}
\end{figure}

IRDC G53.2, seen as several separate filamentary clouds in the mid-IR 
images (Figure~\ref{fig1}), 
coincides with a CO cloud at $v \sim 23$~\kms. 
We use the $^{13}$CO $J=$ 1--0 data from the Boston University$-$Five College Radio 
Astronomy Observatory Galactic Ring Survey (GRS; \citealt{jackson06})
to determine the distance and the mass of IRDC G53.2.
Figure~\ref{fig2} shows the integrated intensity and the mean velocity 
distributions of the $^{13}$CO emission in the area of IRDC G53.2. 
We integrate the GRS cube data from 15 to 30~\kms, 
which covers essentially all of the CO emission associated with the 
cloud (see the average spectrum of the cloud in the inset of Figure~\ref{fig2}).
In Figure~\ref{fig2}, the cloud is elongated over $\sim$1$^{\circ}.5$ 
long in the Galactic plane (for the cloud size determination, see below),
but we note that the CO cloud at low brightness level extends as far 
as $\gtrsim$2$^{\circ}.4$.
The mean velocity of the cloud varies from 22 to 24~\kms\ from 
the western to the eastern part of the cloud (Figure~\ref{fig2}), 
and the mean velocity of the entire cloud is +22.9~\kms. 
Assuming a flat rotation curve with $R_{\sun}=8.5$~kpc 
and $\Theta_{\sun}=220$~\kms, 
the kinematic distance to the cloud is 1.7 kpc, and 
we adopt this as the distance to IRDC G53.2.\footnote{For comparison, 
\citet{ragan14} identified the same molecular cloud 
in their study of filamentary molecular clouds (F54.0$-$52.0 in their Table 1) and   
assigned a kinematic distance of 2.0 kpc using the Galactic rotation model 
of \citet{reid09}, which is based on high-mass star-forming regions.}

In order to obtain the mass of IRDC G53.2, we first derive the $^{13}$CO column 
density (\nco) map. In principle, the integrated intensity map in Figure 2 
can be converted to an \nco\ map if the excitation temperature (\tex) is known
under the LTE assumption \citep[e.g.,][]{rohlfs00}.
One may assume that \tex\ is 
equal to the observed brightness temperature of 
the $^{12}$CO $J=$~1--0 emission, which is usually optically thick, but there are 
uncertainties associated with this method such as unknown beam-filling factors
or pointing/calibration errors from different observations.
In addition, 
the available $^{12}$CO $J=$~1--0 data from the Massachusetts$-$Stony Brook (UMSB) 
Galactic Plane Survey \citep{clemens86,sanders86} are rather poorly sampled. 
We therefore instead adopt $T_{\rm ex}=10$~K which is a typical excitation 
temperature of IRDCs determined from CO observations \citep{heyer09, roman10,liu14}. 
(The median \tex\ from these works might be somewhat below 10 K, e.g., 8--9 K.)
We note that, however,
$T_{\rm ex}=10$ K may be too low for some regions in IRDC G53.2.
The main-beam brightness temperature (\tmb) 
of the $^{13}$CO $J=$~1--0 line is given as 
$T_{\rm mb}=(J_\nu(T_{\rm ex})-J_\nu(T_{\rm bg}))(1-\exp(-\tau_{\rm 13}))$,
where $J_\nu(T)=5.29/(\exp(5.29/T)-1)$, $T_{\rm bg}=2.73$~K, 
and $\tau_{\rm 13}$ is the optical depth of $^{13}$CO $J=$~1--0 
emission \citep[e.g.,][]{rohlfs00}.
Assuming $T_{\rm ex}=10$ K and $\tau_{\rm 13}=3$, which is a typical 
optical depth for most IRDCs \citep{liu14}, the maximum 
allowed \tmb\ of the $^{13}$CO $J=$~1--0 line is 6.4~K, whereas toward 
the cloud cores (e.g., areas around 1.2 mm contours in Figure~\ref{fig1}) 
\tmb\ is higher than this, in some regions as high as 20 K.
The excitation temperature in these areas also must be high, and 
the $^{13}$CO emission might be 
optically thick, which is not unusual for IRDCs \citep[e.g.,][]{liu14}. 
We indeed confirmed that the $^{13}$CO line intensities are comparable 
to the $^{12}$CO line intensities in these areas from the UMSB data.  
Therefore, 
we derive the \nco\ map of IRDC G53.2 by 
applying different assumptions for inner and outer areas
of the cloud. For the outer areas with $T_{\rm mb}\le 6.4$~K, 
we assume $T_{\rm ex}=10$~K, while for the inner areas 
with $T_{\rm mb}\ge 6.4$~K, instead of assuming \tex, we assume a constant 
optical depth of $\tau_{\rm 13}=3$.
This is an ad hoc method but will provide a smooth and 
reasonable \nco\ map until a more accurate map is obtained from 
detailed molecular line studies. (The resulting \nco\ map is found in 
Figure~\ref{fig9}.)

The \nco\ map is then converted to a mass density map 
assuming $^{12}{\rm C}/^{13}{\rm C}=60$ \citep[][their Equation (3)]{milam2005} and 
$n(^{12}{\rm CO})/n({\rm H}_2)=1.1\times 10^{-4}$ \citep{pineda10}.
The total mass of the cloud depends on how we fix its outer boundary. 
For the analysis of the star formation activity in this study, we define 
the boundary of IRDC G53.2 by the $^{13}$CO integrated 
intensity contour at $\int T_{\rm mb} dv=6.3$~K \kms\ (Figure~\ref{fig1}), 
which corresponds to $N(^{13}{\rm CO}) \sim 1\times 10^{16}$~cm$^{-2}$.
This boundary is chosen such that the IRDC includes the dark filamentary 
features seen in mid-IR (24~\um) but excludes bright emission nebulosities 
at 8 and 24~\um, which are likely foreground regions, to avoid 
contamination from unrelated mid-IR sources as much as possible.
Adopting the boundary at 
$N(^{13}{\rm CO}) = 1\times 10^{16}$~cm$^{-2}$,
the cloud mass is $6.2\times 10^4 M_{\sun}$, with a mean 
mass surface density of $0.076$ g~cm$^{-2}$.
The cloud, however, has a low-density envelope (Figure~\ref{fig2}), 
and we account for this by adopting 
$N(^{13}{\rm CO}) = 3\times 10^{15}$~cm$^{-2}$ (the $^{13}$CO 
integrated intensity contour at $\int T_{\rm mb} dv\sim2.1$~K~\kms) 
or $A_{V}=1$ mag as the cloud boundary,
which is slightly beyond the outermost contour in Figure~\ref{fig2}.
This yields an entire cloud mass of $1.0\times 10^5 M_{\sun}$ and 
a mean surface density of $0.033$ g~cm$^{-2}$.\footnote{\citet{ragan14} 
obtained $6.8\times 10^4 M_{\sun}$ as the mass of the molecular cloud.}
The derived parameters of IRDC G53.2 are listed in Table~\ref{tbl1}.

\section{PSC of IRDC G53.2}

We make a PSC of IRDC G53.2 region to search for
YSO candidates. We use two {\it Spitzer} inner Galactic plane 
surveys as a major data set in this study:
GLIMPSE and MIPS GALactic plane survey (MIPSGAL; \citealt{carey09}).
These {\it Spitzer} legacy programs cover the inner Galactic plane 
at the four IRAC bands of 3.6, 4.5, 5.8, and 8.0~\um\ (GLIMPSE) 
and the two MIPS bands of 24 and 70~\um\ (MIPSGAL). 
Other published catalogs such as the Two Micron All Sky Survey (2MASS) 
PSC \citep{2mass}\footnote{http://www.ipac.caltech.edu/2mass/releases/allsky} 
and {\it MSX} PSC \citep{egan03}\footnote{http://heasarc.gsfc.nasa.gov/W3Browse/all/msxpsc.html}
are also used as complements. 

\subsection{Point-source Photometry of MIPS 24~\um\ Mosaics}

We perform point-source photometry of MIPSGAL 
24~\um\ data to extract point sources in IRDC G53.2
since our first criterion to build a PSC of 
the IRDC G53.2 area is the presence of 24~\um\ emission.
We start with four 24~\um\ mosaics centered at 
$(l, b)=$ $(53^{\circ}.0, 0^{\circ}.5)$, $(53^{\circ}.0, -0^{\circ}.5)$, 
$(54^{\circ}.0, 0^{\circ}.5)$, and $(54^{\circ}.0, -0^{\circ}.5)$,
which are distributed by MIPSGAL v3.0 Data Delivery.\footnote{http://irsa.ipac.caltech.edu/data/SPITZER/MIPSGAL}
The size of each mosaic is 1$^{\circ}.1 \times 1^{\circ}.1$, and 
the pixel size is 1$\arcsec$.25.
We use the Astronomical Point source EXtractor (APEX) software 
in the MOsaicking and Point-source EXtraction (MOPEX) package 
developed at the {\it Spitzer} Science Center%
\footnote{http://irsa.ipac.caltech.edu/data/SPITZER/docs/dataanalysistools/tools/mopex}
to extract point sources and compute photometry by 
the point response function (PRF) fitting method based on 
Recipe 24 of the {\it Spitzer} Data Analysis Cookbook.%
\footnote{http://irsa.ipac.caltech.edu/data/SPITZER/docs/dataanalysistools/cookbook}
We first extract bright sources in the background-subtracted mosaics
generated by taking a median filter of the images 
with an 11~pixel by 11~pixel filter and remove the Airy rings 
around the bright sources so that the Airy rings are not detected 
as faint objects; subsequently, we extract faint sources
from the above images.
We measure the flux of the extracted sources and scale them, following 
the Cookbook, by a factor of 1.46 to correct lost fluxes 
due to a small background filter size.
The magnitude zero point we use to convert the estimated fluxes to magnitudes 
is 7.14~Jy as reported in the MIPS Data Handbook v.3.2.1.

The number of sources within the boundary of IRDC G53.2 
presented in Figure~\ref{fig1} from the above procedure is 844.
Here we exclude the sources within the GRS $^{13}$CO contour at 
$(l, b) \sim (53^{\arcdeg}.2, 0^{\arcdeg}.2)$ in Figure~\ref{fig1}.
Although the contour level is the same in the integrated intensity map, 
its lower ($\lesssim$20~\kms) velocity compared to the other region 
of the IRDC (see Figure~\ref{fig2}) and bright emission in 24~\um\ indicate 
that the sources in this region are unlikely associated with IRDC G53.2.
Among the 844 sources in the IRDC, the faintest source is 8.41 mag, and 
the magnitude from which 
the number of sources drops is $\sim$7.8 mag. 
Since the background emission around the IRDC region is 
bright and complicated, we do not extract fainter sources 
for reliability.

\subsection{Merging with GLIMPSE Catalog}

The 24~\um\ sources from the PRF photometry are then matched to 
the GLIMPSEI v2.0 Catalog and Archive with a matching radius 
of 2$\arcsec$ to make a PSC of the IRDC G53.2 area.
As mentioned above, the 24~\um\ mosaics have complicated background
emissions, so residual emissions in the background-subtracted mosaics 
are sometimes detected as a source during the point-source extraction process. 
After matching the sources, we remove the false ones
by visual inspection comparing the original and background-subtracted mosaics.
Finally, 369 sources out of 844 are matched with the GLIMPSEI 
Catalog or Archive.
Table~\ref{tbl2} presents the coordinates and IR magnitudes of 
the 369 matched sources. 
The number of sources whose IRAC magnitudes are from the GLIMPSEI Archive 
is 51, and we put a superscript {\it ``a''} to their classes 
in column 13 (see Section~4 for the source classification).
From the PRF photometry by APEX, we only get uncertainties from the fitting, 
whereas there is an absolute uncertainty (4\% for 24~\um) 
from the pipeline calibration according to MIPS Instrument Handbook v.3.%
\footnote{http://irsa.ipac.caltech.edu/data/SPITZER/docs/mips/mipsinstrumenthandbook}
Therefore, we add both in quadrature and present in Table~\ref{tbl2}.
Note that the 24~\um\ errors in Table~\ref{tbl2} are underestimated 
because they do not account for uncertainties from other components such as 
background variations.
Since we choose completeness rather than reliability, 
we do not make any criteria for using the GLIMPSEI Catalog/Archive. 
The median uncertainty of the IRAC-band magnitudes of the sources 
in Table~\ref{tbl2} is 0.053 mag, 
and $\lesssim 9\%$ of the sources have any one of the IRAC-band magnitudes 
with its error $\gtrsim$0.2~mag. 

Because of strong and complicated background emission, 
many of the unmatched sources are probably false detections 
arising from the residuals after removing the Airy rings or 
from extended emission. There may be genuine sources as well 
that are very deeply embedded, e.g., Class 0 YSOs.
Those not detected in the IRAC bands, however, are beyond
the scope of this study, so we do not further examine 
the unmatched sources here.
We note that one of the unmatched sources 
at (R.A., decl.) = (19:29:16.18, +17:56:10.32)
is very bright at 24~\um, with its magnitude of 1.04~mag,
but not detected in any IRAC bands. 
Our new recent observation in near-IR using adaptive optics reveals 
that this source is, in fact, composed of two stars separated 
by $\sim$0$\arcsec$.75 (H.-J.Kim et al., in preparation).
We measure its IRAC band fluxes from aperture photometry, 
but extended features, particularly remarkable at 5.8 and 8.0~\um,
make the measured fluxes less reliable; therefore, 
we exclude this source from our catalog.

\subsection{Final Point-source Catalog}

From the MIPS 24~\um\ mosaics and GLIMPSE Catalog/Archive, 
369 sources in total are identified in IRDC G53.2 
and listed in Table~\ref{tbl2}. 
Among them, 302 sources are also identified in 
2MASS PSC, so we include their 2MASS 
magnitudes in Table~\ref{tbl2} (columns 4--6).

There are four saturated sources in the MIPS 24~\um\ mosaics 
that are not extracted by APEX. 
For these sources, we use MSX PSC as a complement 
to their IRAC 8~\um\ and MIPS 24~\um\ magnitudes.
The 2MASS and IR magnitudes of these four sources with 
their coordinates from the 2MASS PSC are presented in Table~\ref{tbl3}.
Note that No.~2 is so close to No.~1 that it is not resolved in 
the data from {\it MSX}, which has a larger beam size than {\it Spitzer}. 

\section{Selection and Classification of YSO Candidates}

We select and classify YSO candidates in our PSC (Table~\ref{tbl2}) 
based on the IR spectral index defined as
$\alpha = d~{\rm log}\,(\lambda\,F_{\lambda})/d~{\rm log}\,(\lambda)$ \citep{lada87}.
Spectral index can be used to quantify the slope of spectral energy distributions (SEDs) 
of YSOs in the IR and to divide YSOs into several classes.
We compute the spectral index of each object in Table~\ref{tbl2} by 
least-squares linear fitting between 2 and 24~\um.
Among 369 sources, we only fit 347 sources 
that are detected in more than three IRAC bands to make enough data points 
and classify them as follows \citep{greene94,billot10}:
$0.3 \leq \alpha$ as Class I, the youngest evolutionary class 
whose SED is rising toward mid-IR, which indicates 
the presence of a dusty envelope infalling onto a central protostar;
$-1.6 \leq \alpha < -0.3$ as Class II, which are pre-main-sequence stars 
with warm optically thick dusty disks;
$-2.56 \leq \alpha < -1.6$ as Class III, objects whose SED is 
mostly photospheric emission in near-IR but which show some excess 
emission at longer ($>$20~\um) wavelengths;
and $\alpha < -2.56$ as photospheric-emission-only (photospheric, hereafter) sources.
We also include flat-spectrum ($-0.3 \leq \alpha < 0.3$), 
whose evolutionary status is likely between Class I and II, 
although flat-spectrum sources could be considered  as 
Class I \citep[e.g.,][]{calvet94}.
The computed spectral indices $\alpha$ with 1$\sigma$ errors and 
the results of classification are presented in 
Table~\ref{tbl2} (columns 12 and 13).
We also compute the spectral indices of the MIPS-saturated sources 
using available data and present their $\alpha$ and classes 
in Table~\ref{tbl3} (columns 12 and 13).

As a next step, we examine the sources that are excluded 
from the fitting owing to insufficient data. 
If a source is bright in 24~\um\ but not detected in shorter wavebands, 
it is possible that the source is deeply embedded.
In the investigation of the young, embedded cluster NGC 1333,
\citet{gutermuth08} suggested that 
any source that lacks detections in some IRAC bands but is bright 
at MIPS 24~\um\ ($[24] <  7$ and $[X] - [24] > 4.5$ mag, where $[X]$ is 
the photometry for any IRAC detection available) is likely 
a deeply embedded protostar.
Following this, we check 22 sources detected in only one or two IRAC bands and 
find that 16 sources satisfy the criterion above.
Nondetection of these deeply embedded protostars in the IRAC bands may indicate
that they are Class 0 protostars, which are very young accreting protostars 
in an earlier evolutionary phase than Class I \citep{andre02}.
However, Class 0 protostars are not distinguishable from Class I objects 
by using $\alpha$ \citep{evans09}, and 
Class 0 protostars are originally defined by submillimeter flux \citep{andre02}.
Since we concentrate on mid-IR properties of YSOs, 
an exact classification of Class 0 protostars is not necessary in this study;
therefore, we include the deeply embedded protostars in Class I 
and put a superscript {\it ``d''} to their classes in Table~\ref{tbl2}. 
The remaining six sources that do not meet the criterion for deeply embedded protostars 
are classified as `No Class' sources in the table.
 
Finally, among 373 sources in IRDC G53.2,
including the saturated sources (two Class I and two Class II objects) 
in the MIPS 24~\um\ (Tables~\ref{tbl2} and \ref{tbl3}),
we find 78 Class I (21\%), 66 flat-spectrum (18\%), 135 Class II (36\%), 
62 Class III (17\%), and 26 photospheric (7\%) sources, 
leaving 6 ($\sim 1\%$) sources without classification.
In further sections, we refer to Class I, flat-spectrum, Class II, and Class III 
objects as YSO candidates.

\section{ Contaminations}
\subsection{Extragalactic Contamination}

\begin{figure}
\begin{center}
\includegraphics[width=0.4\textwidth]{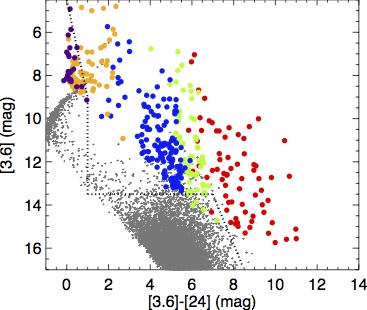}
\caption{[3.6] vs. [3.6]--[24] plot for objects in the 
IRDC G53.2 region and SWIRE \citep{lonsdale03} ELAIS N-1.
Red, green, blue, yellow, and purple circles indicate 
Class I, flat-spectrum, Class II, Class III, and photospheric 
sources, respectively. Gray dots are SWIRE ELAIS N-1 samples.
The dotted line divides the regions occupied mostly by SWIRE-type 
contaminants (galaxies and diskless stars) and where YSOs \citep{rebull11}.
\label{fig3}}
\end{center}
\end{figure}

MIPS sources generally contain extragalactic backgrounds 
as well as Galactic sources, but 
we expect most sources in IRDC G53.2 to be Galactic 
since the IRDC is in the Galactic plane ($b\sim0^{\circ}$).
Although both galaxies and YSOs show IR-excess emission,
they can be distinguished in a [3.6] versus [3.6]$-$[24] plot 
\citep{rebull11}.
In the [3.6] versus [3.6]$-$[24] plot, \citet{rebull11} compared 
the extragalactic sources from the {\it Spitzer} Wide-area Infrared
Extragalactic Survey (SWIRE; \citealt{lonsdale03}) 
European Large Area {\it ISO} Survey (ELAIS) N-1 extragalactic field 
with the YSO candidates in North American and Pelican Nebulae (NAN)
and Serpens (see their Figure~10), and
they found that the YSO candidates occupy a different region 
from the SWIRE-type contaminants, which include galaxies and 
diskless stars.
We also compare the IRDC G53.2 sources (Tables~\ref{tbl2} and \ref{tbl3})
with the SWIRE ELAIS N-1 samples%
\footnote{SWIRE ELAIS N1 Region 24~\um\ Spring '05 Spitzer Catalog from IPAC Infrared Science Archive (http://irsa.ipac.caltech.edu/applications/Gator)}
in Figure~\ref{fig3}.
In the figure, 
the distribution of the IRDC 53.2 sources (colored circles) is distinctively 
separated from the SWIRE-field samples (gray dots)
except for the ones with mostly stellar photospheric emission 
(i.e., Class III and photospheric sources),
which have [3.6]$-$[24]$\sim$0.
The location of the YSO candidates in IRDC G53.2 
in the [3.6] versus [3.6]$-$[24] plot is similar to those of the YSO candidates 
in NAN and Serpens as well \citep[Figure~10 of][]{rebull11}.
In addition, we searched the SIMBAD database\footnote{http://simbad.u-strasbg.fr/simbad/} 
and found no counterpart for each of our 373 sources 
among known extragalactic sources within a 5$\arcsec$ radius.
We therefore conclude that essentially all of our sources are likely of 
Galactic origin, and there is very little contamination from extragalactic sources.

\subsection{Field Star Contamination}

\begin{figure}
\begin{center}
\includegraphics[width=0.45\textwidth]{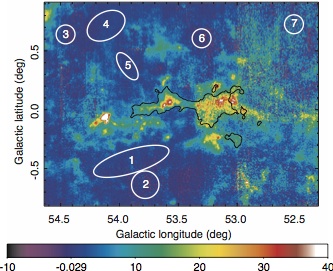}
\caption{
Same as the left panel of Figure~\ref{fig2}, but 
the integration has been done over the entire velocity range 
of the GRS survey, i.e., $\vlsr=-5$ to +85~\kms.
Seven white ellipses and circles with numbers are control fields, 
and the black contour shows the boundary of IRDC G53.2.
The scale bar indicates the integrated intensity scale.
\label{fig4}}
\end{center}
\end{figure}

While we do not expect extragalactic contamination toward the Galactic plane, 
there is nonetheless contamination by foreground/background stars.
In order to account for the amount of contamination from these field stars,
we analyze control fields and compare the population of point sources 
between the control fields and IRDC G53.2.
We select seven control fields around IRDC G53.2 
where there are no CO clouds in the GRS $^{13}$CO $J=$~1--0 data.
The control fields in the GRS integrated intensity map are presented in
Figure~\ref{fig4} by white ellipses and circles with their numbers.
In the figure, the integration of the GRS cube data has been done over the 
entire velocity range of the GRS survey, i.e., $\vlsr=-5$ to +85~\kms, 
and the scale bar indicates the integrated intensity scale.
The boundary of IRDC G53.2 is shown by a black contour as well.

We identify point sources in the control fields from the MIPS 24~\um\ mosaic, 
match them to the GLIMPSEI Catalog/Archive, 
and perform the same spectral index analysis 
as we did for the sources in IRDC G53.2.
In the seven control fields, 218 sources are classified in total.
Table~\ref{tbl4} presents the results of the control field analysis and 
the comparison to IRDC G53.2.
Among the sources in the control fields (column 2), 
about 90\% are Class III or photospheric sources 
with little excess emission in IR.
The others classified as Class I, flat-spectrum, or Class II show similar IR excess 
to the YSO candidates. 
These IR-excess sources are, however, not likely YSOs but the sources 
with similar colors to YSOs such as AGB stars or planetary nebulae 
because they are mostly isolated and placed where there are no molecular clouds.
Since the area of the IRDC is $\lesssim$0.5 times of 
the total area of the control fields,
we scale the number of sources in the control fields to the IRDC G53.2 area 
to compute the expected numbers of each class (column 3) and 
compare them with the numbers of the sources in the IRDC (column 4).

The comparison between the control fields and IRDC G53.2 implies that
there is a negligible contamination from field stars 
in earlier classes of YSO candidates (i.e., Class I, flat-spectrum, and Class II), 
while about 50\% of field star contamination exists among Class III objects.
In addition, all photospheric sources identified 
in the IRDC G53.2 area are not likely the genuine members of the IRDC 
but foreground/background stars.
We note that we detect fewer ($\lesssim$50\%) photospheric sources 
in IRDC G53.2 than expected from the control field analysis.
Photospheric sources without strong IR excess are mostly faint in 24~\um;
for example,
the median [24] magnitudes of each class in IRDC G53.2 are 
4.5, 6.2, 6.8, 7.1, and 7.7 for Class I, flat-spectrum, Class II, Class III, 
and photospheric objects, respectively.
Since the extinction toward IRDC G53.2 is higher than the control fields
without molecular clouds, we probably have failed to detect a large number of 
photospheric sources owing to extinction, which results in a smaller number of 
the photospheric sources in the IRDC compared to the scaled number from 
the control fields.
Here, we only compare the sources with the MIPS 24~\um\ photometry (Table~\ref{tbl2})
because there are no saturated sources in the MIPS 24~\um\ mosaic
in the control regions. However, even if we include the MIPS-saturated sources 
that are presented in the parentheses in column 4 of Table~\ref{tbl4}, 
its small number does not affect the results.

Based on the comparison with the control fields, we expect that our 
PSC of IRDC G53.2 is contaminated 
with 1 Class I, 3 flat-spectrum, 6 Class II, and 29 Class III objects.  All of the 
photospheric sources in our catalog are likely to be contaminants.  
If we removed this field star contamination, we would be left with 308 sources 
in total (including four MIPS-saturated sources), and the final census 
of YSO candidates would be 77 Class I (25\%), 63 flat-spectrum (21\%), 
129 Class II (42\%), 33 Class III (11\%) objects, and 6 ‘No Class’ (∼1\%) sources.

\section{Discussion}

\subsection{Classification Schemes of YSOs}

\subsubsection{Based on Spectral Index, $\alpha$}
In classification of YSOs, 
an empirical scheme based on the slope of SEDs 
was first constructed by \cite{lada84} 
and more quantitatively developed by \cite{lada87} adopting
the definition of spectral index $\alpha$.
Since then, \citet{greene94}, introducing flat-spectrum (see Section~3.1), 
has developed the classification system 
in the study of the $\rho$ Oph YSO population.
Whereas the original definition of $\alpha$ used wavelengths 
between 2 and 20~\um\ \citep{lada84}, 
several recent studies have used only the four IRAC bands 
to compute $\alpha$~\citep[e.g.,][]{lada06,billot10}.
Regarding differences in the choice of wavelength range, 
\citet{greene94} claimed no deviation between $\alpha$ 
computed using 2.2--20~\um\ and 2.2--10~\um, 
but \cite{evans09} classified YSOs in five nearby molecular clouds 
in the c2d project \citep[][hereafter the c2d clouds]{evans03} using 
both 2--24~\um\ and the IRAC bands and showed that 
using only the IRAC bands moves 
sources from earlier to later classes, resulting in 10\%--15\% difference 
in inferred lifetimes for the earlier SED phases.

\begin{figure}
\begin{center}
\includegraphics[width=0.45\textwidth]{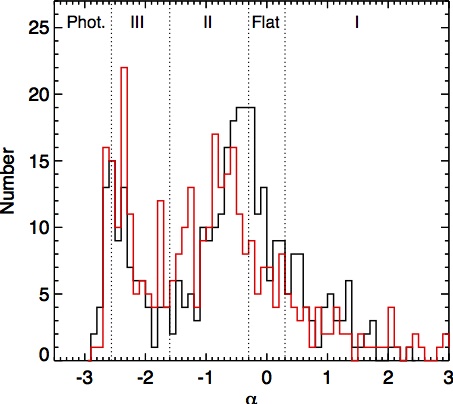}
\caption{Histogram of spectral indices of the point sources in IRDC G53.2. 
Black and red lines are spectral indices computed by least-squares linear fitting 
in the range 2--24~\um\ and only using the IRAC bands, respectively. 
Dotted lines divide the regions for each class.
\label{fig5}}
\end{center}
\end{figure}

\begin{figure}
\begin{center}
\includegraphics[width=0.45\textwidth]{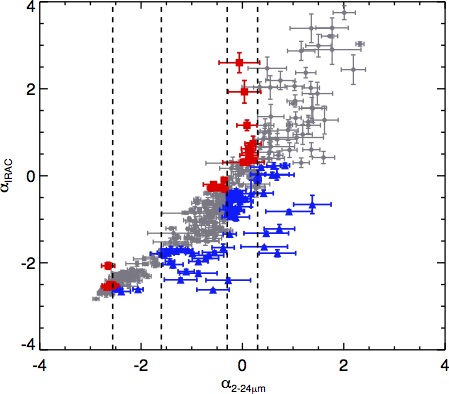}
\caption{Comparison between spectral indices of the point sources 
in IRDC G53.2 computed using 2--24~\um\ (\aall) 
and using only the IRAC bands (\airac).
Gray circles are the objects whose classes from the both spectral indices are the same.
Blue triangles are the objects whose class from \airac\ is later than that 
from \aall, and red squares are the objects whose class from \airac\
is earlier than that from \aall.
Dashed lines divide the regions for each class based on \aall.
\label{fig6}}
\end{center}
\end{figure}

In Section~4, we classified the point sources in IRDC G53.2 by 
least-squares linear fitting between 2 and 24~\um. 
In order to examine the difference in classification depending on 
the use of wavelength range, we classify the sources again 
using only the IRAC bands and compare the results to the one 
we have from 2--24~\um\ in Figure~\ref{fig5}.
In Figure~\ref{fig5}, 
the histogram of spectral index from only using the IRAC bands, drawn 
in red, shows a relatively lower number of flat-spectrum and higher number 
of Class III than the histogram of spectral index from 2--24~\um, drawn 
in black. 
When using only the IRAC bands, 
we get fewer Class I (by $\sim$7\%), fewer flat-spectrum (by $\sim$40\%),
comparable Class II, and more Class III (by $\sim$43\%).
Since a direct comparison in numbers of each class does not account for
uncertainties from fitting in the computation of spectral index,
we show a spectral index from 2--24~\um\ (\aall) versus 
spectral index from the IRAC bands (\airac) plot with 
their 1$\sigma$ uncertainties from fitting in Figure~\ref{fig6}.
As shown in Figure~\ref{fig6}, 
most sources plotted by gray circles have consistent classes
from the both spectral indices.
Red squares in Figure~\ref{fig6} are the sources that are classified 
as earlier class by \airac\ with its fraction of $\sim$7\%, but 
they fall into the same classes by \aall\ 
when accounting for uncertainty.
Blue triangles in the figure are the sources that are classified 
as later class by \airac\ with its fraction of $\sim$20\%; however,
if we account for the uncertainties, the fraction of sources classified by
later class by \airac\ is 12\%. 
From the above comparison, 
the uncertainty in YSO candidate classification in this study 
from the choice of wavelength range in determination of spectral index 
will be $\sim$12\%.

\subsubsection{Based on Mid-infrared Colors}
YSOs are often classified based on mid-IR colors.
For example, \citet{gutermuth08,gutermuth09} established a mid-IR 
color-based method that can robustly distinguish YSOs and mitigate
the effects of contamination and reddening. Their mid-IR color-based 
method primarily uses the IRAC magnitudes/colors and additionally 
uses 24~\um\ detection for further classifying ``Transition Disk'' Class II 
sources, deeply embedded protostars, and highly reddened Class II sources. 
The color-based classification of YSOs in principle relies on the mid-IR 
excess emission, as the spectral-index-based classification does, but 
different criteria between the two may give different classification. 
In order to check this possibility,
we apply the classification scheme described in \citet{gutermuth09} to 
the sources in IRDC G53.2 and compare the results between 
the color-based and the \aall-based classification.
Among 369 sources in Table~\ref{tbl2} (excluding the saturated sources 
in 24~\um), 70 and 189 sources are classified as Class I and Class II, 
respectively, based on mid-IR colors, where Class I includes 16 deeply 
embedded protostars and Class II includes 52 transition disk sources 
and 7 highly reddened Class II sources. The others are the sources with 
photospheric colors or the sources that lack the IRAC magnitudes.
Comparing the number of objects in each class, we find that 
most Class I ($\gtrsim$80\%), Class II ($>$90\%), and 
Class III/photospheric sources ($\sim$100\%) from the \aall-based 
classification  are well consistent with the color-based classes. 
In the case of flat-spectrum sources, however, $\sim$80\% are 
classified as Class II by the color-based classification, while they are 
supposed to be included in Class I according to \citet{gutermuth09}.

A possible reason for this discrepancy is an ambiguity of flat-spectrum. 
Since flat-spectrum is an evolutionary stage between Class I 
and Class II, separating flat-spectrum sources from either class
will be highly uncertain. 
The wavelength range used in classification can be a reason as well.
Although the color-based classification is made from IRAC to 
MIPS 24~\um\ wavelengths, classes are mainly determined 
by IRAC colors because the 24~\um\ magnitudes are only 
additionally used to classify anomalies such as transition disk 
or highly reddened Class II.
Therefore, a source with an SED slope in the IRAC wavelengths 
close to the boundary between flat-spectrum and Class II
but bright in 24~\um\ may be classified as Class II by 
the color-based scheme, whereas it would be classified 
as flat-spectrum (i.e., Class I) by the \aall-based scheme. 
This trend is also seen in the comparison between the \aall-based 
and \airac-based classification discussed above (Section~6.1.1). 
We note that extinction may be an another factor that moves 
flat-spectrum to Class II, and it will be discussed below in detail.
In summary, the mid-IR color-based classification of
YSOs is consistent with the \aall-based classification scheme in general, 
except when classifying flat-spectrum sources. In such objects, there is some 
ambiguity between Class I and Class II sources, which is likely dependent 
on the wavelength ranges used.
The difference between the two classifications will affect the number ratio 
of Class II to Class I objects by a factor of two or three (Section~6.2.1).

\subsubsection{Based on Physical Properties of SEDs}
\begin{figure}
\begin{center}
\includegraphics[width=0.4\textwidth]{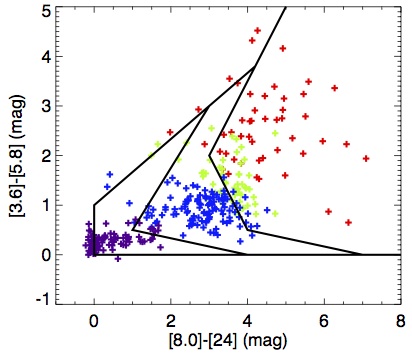}
\caption{Color-color diagram using the IRAC bands at 3.6, 5.8, 8.0~\um\ 
and the MIPS 24~\um. Red, green, blue, and purple crosses represent 
Class I, flat-spectrum, Class II, and Class III/photospheric sources 
in IRDC G53.2, respectively, classified based on the spectral index \aall. 
Black solid lines divide the areas filled by Stage I, II, and III 
sources from right to left, adopted from Figure~18 of \citet{robitaille06}.
\label{fig7}}
\end{center}
\end{figure}

On the other hand, \citet{robitaille06} adopted a ``Stage'' classification.
Stage is based on the physical properties of SEDs, thus 
referring to the actual evolutionary stage rather than Class, 
which is only based on the slope of SEDs.
Although Stage 0/I/II/III is analogous to Class 0/I/II/III, 
it is not a one-to-one correspondence because of the inclination effect. 
For example, pre-main-sequence stars with an ``edge-on'' disk 
can be classified as flat-spectrum or Class I by their SED slopes, 
and \citet{crapsi08}, in their study on the nature of embedded YSOs 
using radiative transfer modeling, claimed that
34\% of the Stage II sources could be misclassified as flat-spectrum. 
Despite the possible misclassification, however, 
the classes of the YSO candidates in IRDC G53.2 
classified based on spectral index \aall\ agree fairly well with 
the stages in a color-color diagram as presented in Figure~\ref{fig7}.
In Figure~\ref{fig7}, Class I, II, and III objects 
mostly fall into the areas of Stage I, II, and III \citep{robitaille06}, 
respectively.
Flat-spectrum sources again show an ambiguity between 
Stages I and II, but $\sim$65\% are located in the Stage I area.
The others placed in the Stage II area may represent 
the ones with large inclination angle.
Comparison of ``Class'' and ``Stage'' classification from the
[3.6]$-$[5.8] versus [8.0]$-$[24] color-color diagram 
indicates a general agreement between the two, 
although there are some exceptions.
Detailed investigation on the physical properties 
of the YSO candidates in IRDC G53.2 based on 
the analysis of the individual SEDs will be addressed 
in our forthcoming paper.  

\subsubsection{Effects from Extinction on Classification}
Embedded star-forming regions in a dense molecular cloud 
suffer high extinction from dust and reddening. 
High column density, and thus high extinction, toward IRDC G53.2
may affect the classification of YSOs based on SED shapes,
but it may not be significant because of 
much smaller extinction in mid-IR than in optical.
Previous studies using {\it Spitzer} data also show that 
the effects from extinction on the overall YSO classification 
are small.
\citet{evans09} compared spectral index of YSOs in the c2d clouds
before and after extinction correction and found that 
the effect from extinction was small (10\%--20\%), probably smaller 
than other sources of uncertainty. 
\citet{billot10} also found that the YSO census in the Vulpecula 
OB association (hereafter Vul OB1) based on the mid-IR color 
method showed only a marginal difference 
after they corrected for extinction. 
This indicates that classification schemes based on mid-IR excess 
are only moderately affected by extinction. However, it is also possible 
that extinction effects in our analysis of the IRDC G53.2 YSO population, 
which is much further away and embedded in a denser molecular cloud,
may not be as negligible as in the above studies, and 
we will briefly discuss expected effects from extinction 
on the YSO census in IRDC G53.2 below.

The local extinction in star-forming regions may be derived using 
the near-IR colors of stars observed in each region and comparing these 
with their intrinsic colors \citep[e.g.,][]{evans09,gutermuth09}. However, 
since this method is limited to nearby regions, we cannot apply it to 
IRDC G53.2. Extinction in IRDC G53.2 is also not uniform; 
it is larger in the central parts of the molecular cloud, where 
the $^{13}$CO integrated intensity map peaks in Figure~\ref{fig2} (left). 
To investigate the effects of extinction on our YSO census, 
we therefore consider the extreme case of one of the clumps 
in IRDC G53.2 \citep[from][]{butler12} and examine how it could 
possibly affect the YSO classes. 
\citet{butler12} estimated physical properties of starless and early-stage 
cores and clumps in 10 IRDCs, including the one in IRDC G53.2 
\citep[MSX G053.11+00.05 in Figure~\ref{fig1}; core J1 in][]{butler12}
based on the mid-IR extinction mapping technique. 
Their estimation, using the IRAC 8~\um\ images, 
gave a mean mass surface density of $\sim$0.15 g~cm$^{-2}$
or $\tau_{\rm 8\mu m}\sim1.1$ (according to their Equation~(1)) for a clump 
in the {\it MSX} G053.11+00.05. 
From the mid-IR extinction law derived by {\it Spitzer} data 
\citep{flaherty07, chapman09}, we estimate the extinction at 
each IRAC and MIPS 24~\um\ band as follows, although extinction at 
24~\um\ is highly uncertain owing to a small sample size in \citet{flaherty07} 
and \citet{chapman09}: 
$A_{[3.6]}=$ 1.6--1.7~mag, $A_{[4.5]}=$ 1.3--1.4~mag, $A_{[5.8]} \sim$1.2~mag,
$A_{[8.0]} \sim$~1.2~mag, and $A_{[24]} \sim$~1~mag.
If we apply extinction correction to these values in Figure~\ref{fig7}, 
about half of the flat-spectrum sources in the Stage I area 
move to the Stage II area, whereas Class I and Class II sources show 
only a small changes. 
However, we note that this extinction will be close to the upper 
limit in IRDC G53.2. 
The clump in the {\it MSX} G053.11+00.05 is located where the $^{13}$CO 
emission is strong, corresponding to the innermost contour in 
Figure~\ref{fig2}, and extinction of most of the area in the IRDC should be 
much lower. 
For example, the mass surface density (or extinction) 
of the outer part of the clump in the {\it MSX} G053.11+00.05 is $\sim$10\% 
of the mean value \citep[Figure~12 of][]{butler12}, 
which does not make any remarkable changes in Figure~\ref{fig7}.
Therefore, 
extinction in IRDC G53.2 may affect the YSO classification 
in a way to decrease the number of flat-spectrum objects and 
increase the number of Class II objects, but the effect on the overall 
classification is likely insignificant.

\subsection{Comparison of the YSO Population in 
IRDC G53.2 to Other Regions}

While IRDCs are believed to be a probable site of massive star formation,
many of them do not show a signature of active star formation in mid-IR 
represented by 24~\um\ point sources \citep{chambers09}, 
and the clumps in IRDCs are typically not associated with 24~\um\ 
sources \citep{ragan09}. 
In this aspect, IRDC G53.2, with a few hundreds of 
YSO candidates, is a unique place where we can statistically 
investigate the stellar population formed in IRDCs. 
The YSO population in IRDC G53.2 is also able to provide 
the star formation activity occurring in the whole associated molecular cloud 
in which IRDCs are embedded. 
Below, we discuss the stellar population in IRDC G53.2 
by comparing with that of other well-studied star-forming regions 
in different environments (Section~6.2.1). We also compare 
with other IRDCs, some of which are associated with 24~\um\ sources 
and some of which are not (Section~6.2.2), mainly focusing on the evolutionary 
stage of the IRDC. Finally, we estimate the age of IRDC G53.2 
using an analytic model \citep{myers12} based on the YSO census and 
compare it with other nearby star-forming clusters (Section~6.2.3).

\subsubsection{Comparison with Other Star-forming Regions: 
Number Ratio of Class II to Class I}

We first compare the census of YSO candidates in IRDC G53.2 
and other star-forming regions because the number of YSOs in each class 
can provide an estimation of the relative age of a star-forming region,
assuming a constant birthrate \citep{gutermuth09}.
For example, the number ratio of Class II to Class III objects, which can be 
a diagnostic of a fraction of YSOs with disks (i.e., disk fraction), gives 
the age of a star-forming region since disk fraction exponentially decreases
with the age of a star-forming region \citep{mamajek09,ybarra13}.
The use of the ratio of Class II to Class III, however, 
would be inappropriate in the case of IRDC G53.2. 
In IRDC G53.2, the fraction of Class I and flat-spectrum sources 
that may not even form accretion disks is rather high ($\lesssim$50\%), and 
the census of Class III objects and photospheric sources is likely to be incomplete
because of their weak IR-excess emission and faintness at 24~\um\
(see Section~5.2).
Instead, we use the number ratio of Class II to Class I objects 
(\ntwoone, hereafter) to compare the relative age of 
IRDC G53.2 with other star-forming regions. 
As mentioned in Section~4, the evolutionary status of flat-spectrum is rather 
uncertain, and the objects with SEDs of flat-spectrum are often included in Class I 
if they are not separately classified \citep[e.g.,][]{gutermuth09}.
Following this, 
we include flat-spectrum in Class I in the discussion below.

\begin{figure}
\begin{center}$
\begin{array}{cc}
\includegraphics[width=0.4\textwidth]{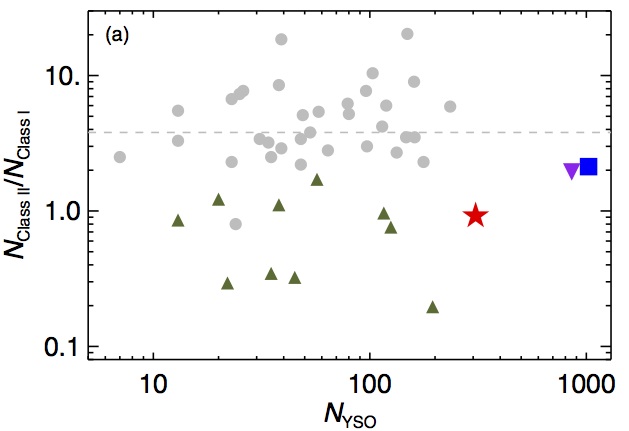} &
\includegraphics[width=0.4\textwidth]{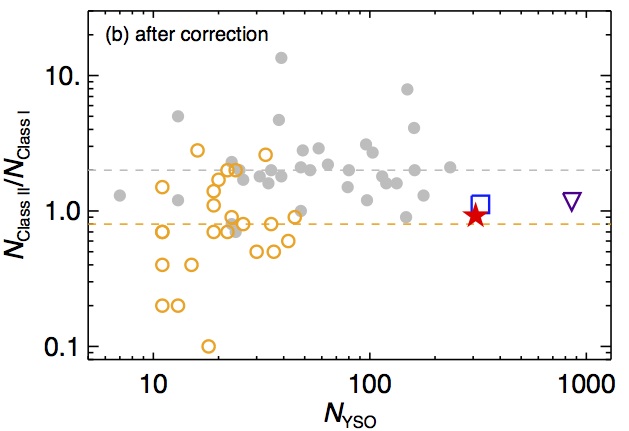}
\end{array}$ 
\end{center}
\caption{(a) Number ratios of Class II to Class I objects vs. the number of YSOs 
of IRDC G53.2 (filled red star) and in other star-forming regions. 
Filled gray circles: 36 nearby star-forming clusters \citep{gutermuth09}
with their median value of 3.7 marked by a gray dashed line;
filled green triangles: filamentary structures in the inner Galactic region 
\citep{bhavya13}; 
filled blue square: the c2d clouds \citep{evans09};
filled purple downward-pointing triangle: Vul OB1 \citep{billot10}. 
(b) Number ratios of Class II to Class I objects after correcting 
for different classification schemes and distances (Section 6.2.1).
Filled gray circles: the clusters in \citet{gutermuth09} adapted to
the \aall-based classification scheme with their median value of 2.0 marked 
by a grey-dashed line;
open orange circles: the clusters in \citet{gutermuth09} after scaling to 
the distance of IRDC G53.2 with the median value of 0.8 marked by 
an orange dashed line; 
open blue square: the c2d clouds after distance correction;
open purple downward-pointing triangle: Vul OB1 based on the \aall-based classification.
\label{fig8}}
\end{figure}

Figure~\ref{fig8}(a) presents \ntwoone\ of IRDC G53.2 
and other star-forming regions from the literature.
In the figure, \ntwoone\ of IRDC G53.2, 
which is 0.9 after removing field star contamination, is marked with 
a filled red star, and the ratios of other star-forming regions from different literature 
are marked with different filled symbols.
Filled gray circles are from a systematic {\it Spitzer} survey 
on 36 young, nearby, star-forming clusters within 1~kpc by \citet{gutermuth09}, 
and a gray dashed line indicates the median value of \ntwoone\ (3.7) 
for these clusters.
\ntwoone\ of the c2d clouds \citep{evans09} and Vul OB1 \citep{billot10} 
are also presented by a filled blue square and filled purple 
downward-pointing triangle, respectively.
Filled green triangles are \ntwoone\ of YSOs that are spatially 
associated with dark filamentary structures in the inner Galactic region of 
$10^{\circ} < l < 15^{\circ}$ and $-1^{\circ} < b < 1^{\circ}$ using 
GLIMPSE data \citep{bhavya13}.
While nearby, low-mass star-forming regions \citep{evans09, gutermuth09} 
and Vul OB1 \citep{billot10} show higher \ntwoone\ than that of 
IRDC G53.2, 
YSOs likely related to IRDCs \citep{bhavya13} show similar or 
lower \ntwoone\ compared to IRDC G53.2.
Direct comparison of the number ratios from different studies, however,
is not appropriate because (1) different classification schemes affect 
the fraction of each class, as discussed in Section 6.1, and 
(2) different distances affect the detection limit so that the number of 
faint sources at 24~\um\ (i.e., later-class YSOs) decreases 
as the distance to the star-forming region increases.

To make an appropriate comparison, we first compute \aall\ of 
YSOs in the star-forming clusters from \citet{gutermuth09} and 
Vul OB1 \citep{billot10} using their flux catalogs and reclassify them. 
\citet{gutermuth09} classified YSOs using their mid-IR colors. 
Since the color-based classification results in fewer Class I 
and more Class II sources than the \aall-based classification (Section~6.1.2), 
\ntwoone\ will become higher if one uses the color-based classification 
scheme. 
After reclassification of their sources based on \aall,
the median of \ntwoone\ for all the clusters becomes smaller, as expected, 
from 3.7 to 2.0. 
New \ntwoone\ distribution and the median value  
are presented in Figure~\ref{fig8}(b) with filled gray circles and 
a gray dashed line, respectively.
\citet{billot10} classified the YSOs in Vul OB1 using \airac, so they 
also have fewer flat-spectrum and more Class II sources and thus 
higher \ntwoone\ (Section 6.1.1 and Figure~\ref{fig5}). 
Our \aall-based classification scheme to the sources in Vul OB1 
decreases \ntwoone\ from 1.9 to 1.2, and the new ratio is marked 
with an open purple downward-pointing triangle in Figure~\ref{fig8}(b). 
For the c2d clouds \citep{evans09}, 
we use their results because they used \aall\ in classification.

Next, we correct for distance by assuming that the nearby regions 
from \citet{evans09} and \citet{gutermuth09} are at the same distance 
as IRDC G53.2 (at 1.7~kpc).  To do this, we scale the fluxes of the YSOs 
and extract the sources brighter than a threshold at 24~\um.
For the threshold, we use 
8.41~mag, the faintest 24~\um\ magnitude in IRDC G53.2.
Larger distance may affect the classification itself as well
owing to higher extinction, but such an effect will be negligible in 
the overall statistics (Section 6.1.4).
We mark distance-corrected \ntwoone\ in Figure~\ref{fig8}(b) 
using open orange circles for the clusters in \citet{gutermuth09} 
with their median value of 0.8 (orange dashed line) 
and an open blue square for the c2d clouds. 
For the clusters in \citet{gutermuth09}, we only consider 
the clusters with total number of YSOs $>$ 10.
In the c2d clouds, $\sim$50\% of Class I sources including 
flat-spectrum remain after distance correction, whereas
only $\sim$26\% of Class II sources remain. 
This indicates that in a region at larger distance, the YSO census 
is likely biased to an earlier class so that \ntwoone\ becomes lower. 
We do not correct distance for Vul OB1 since it is at a similar 
distance (2.3~kpc) to IRDC G53.2.
We summarize the newly computed \ntwoone\ for each region 
after correcting classification and distance in Table~\ref{tbl5}.
Number ratio of Class II to Class I in a star-forming region is 
sensitive to both classification scheme and distance, and 
each factor changes the ratio by a factor of two. 
We do not make any correction for the IRDC-associated YSOs 
from \citet{bhavya13} because their catalog is not available. 
The IRDCs in their samples are mostly farther away (3--6~kpc) 
than IRDC G53.2, so \ntwoone\ will become higher by a factor 
of two if assuming the distance of 1.7 kpc. On the other hand, 
they classified the sources based on \airac, so the use of \aall 
will make \ntwoone\ lower by a factor of two. Therefore, 
their corrected \ntwoone\ may not be very different from the uncorrected 
ones.

Under the same classification criteria and assuming the same distance,
all of the compared regions and IRDC G53.2 show 
similar \ntwoone, which indicates that they are similar in age or 
at a similar evolutionary stage.
We note that there may be more Class I objects in IRDC G53.2 
that are too deeply embedded  to be detected even in 24~\um\ 
because of high column density of the central part of the IRDC. 
Such sources are, however, beyond the scope of this study, 
and further investigation including longer wavebands will be helpful 
to search for them.

\subsubsection{Comparison with Other IRDCs}

Star formation activity in IRDCs has mostly been studied 
on core (10$^{-2}$--10$^{-1}$ pc) or clump (10$^{-1}$--10$^{0}$ pc) 
scales, so a direct comparison of the stellar population in 
different IRDCs, as in Section 6.2.1, is difficult. Some studies, however, 
address the issues on protostars detected at 24~\um\ in the vicinity 
of IRDCs and their association. 
The 24~\um\ point-source detection is one of the signs to trace 
star formation activity in IRDC clumps and related to their evolutionary phase.
IRDC clumps are suggested to evolve from a quiescent clump to 
an active/red clump with an intermediate clump in between.
As clumps evolve, they become warmer and show tracers of 
star formation such as embedded 24~\um\ point sources or 
H$_{2}$O/CH$_{3}$OH maser emission \citep{chambers09,battersby10}.
A large number of YSO candidates detected in 24~\um\ in IRDC G53.2, 
with the associated ``green fuzzies (extended 4.5~\um\ enhancement)" 
and H$_{2}$O/CH$_{3}$OH masers previously found \citep[e.g.,][]{chambers09}, 
indicates that stars are actively forming in IRDC G53.2 and 
IRDC G53.2 is likely at a later evolutionary stage among IRDCs.

\citet{ragan09} investigated stellar content around 11 IRDCs
at 2.4--4.9 kpc using {\it Spitzer} data. 
They found many Class II and a few Class I objects, but 
only $\sim$10\% of them were associated with the IRDC clumps 
identified by the absorption in 8~\um. 
From the clump mass function, which is shallower than the Salpeter 
mass function, and a lack of association between the clumps and 
mid-IR sources, the authors suggested that IRDCs are the precursors 
to stellar clusters in an early phase of fragmentation. 
In the IRDC G011.11-0.12, \citet{henning10} found $\sim$20 embedded 
cores in a diverse mass range (1--240~\msol), 
half of which were associated with the 24~\um\ detection.
From large spacings between the cores, well in excess of 
the Jeans length in the IRDC (see Section 6.3.2 below), 
the authors also concluded that IRDC cores are at an early
stage in protostar formation with a capability of forming massive 
stars and clusters. 
These IRDC clumps in the early phase of fragmentation 
\citep{henning10,ragan09} are very weakly associated with 
24~\um\ point sources with lower surface density of sources in 
mid-IR compared to IRDC G53.2.
This again implies a more evolved status for IRDC G53.2. The cores 
in the IRDC G011.11-0.12 or in the IRDCs in \citet{ragan09} may represent 
an earlier stage than that seen in IRDC G53.2, before active star 
formation has turned on.
Future studies on stellar mass function in IRDC G53.2 or 
on physical properties of the clumps harboring bright 24~\um\ 
sources will be helpful for further comparison between 
IRDC G53.2 and other IRDCs at various evolutionary stages.

\subsubsection{Age Estimation Using Analytic Model}

\citet{myers12} developed an analytic model of protostar mass and 
luminosity evolution in clusters that provides estimates of cluster age, 
protostar birthrate, accretion rate, and mean accretion time, 
under the assumptions of constant protostar birthrate, core-clump accretion, 
and equally likely accretion stopping. 
Based on the model, the age of a star-forming cluster can be described 
as $t \simeq \bar{a}/[\nu(1+\nu^2/2(1-\nu))]$
\citep[Equation (38) of][]{myers12}, where $\bar{a}$ is the accretion timescale 
and $\nu$ is the fraction of protostars among total YSOs in the cluster.
From this, \citet{myers12} estimated the ages and birthrates 
of 31 nearby clusters and complexes using the observed numbers of 
protostars and Class II YSOs 
assuming the accretion timescale of 0.17~Myr
and the equal numbers of Class II and Class III sources
(see Section~5 of \citealt{myers12} for details).
In the above equation, the term $\nu^2/[2(1-\nu)]$ becomes 
negligible as $\nu$ decreases, so that the equation can be approximated 
by \ntwoone\ and the age of a star-forming cluster as 
$t \simeq 2\bar{a}(N_{\rm ClassII}/N_{\rm ClassI} +1/2)$. 
We apply this relation to IRDC G53.2,
assuming the same accretion timescale of 0.17 Myr, 
and the estimated age of IRDC G53.2 is $\sim$0.5 Myr. 
This age may be highly uncertain 
because of several assumptions used in the model, 
particularly the accretion timescale, which varies from 
0.12 to 0.4--0.5~Myr (\citealt{myers12} and references therein).
The age of the nearby star-forming clusters in \citet{gutermuth09} 
using \ntwoone\ after correcting for classification and distance 
(Section~6.2.1) ranges from 0.2 to 1.1~Myr, with a median of 0.5~Myr. 
This is comparable to the estimated age of IRDC G53.2.

\subsection{Spatial Distribution of the YSO Candidates}

\subsubsection{Distribution of the YSO Candidates in 
the Molecular Cloud and Far-IR Emission}

\begin{figure}
\begin{center}
\includegraphics[width=0.7\textwidth]{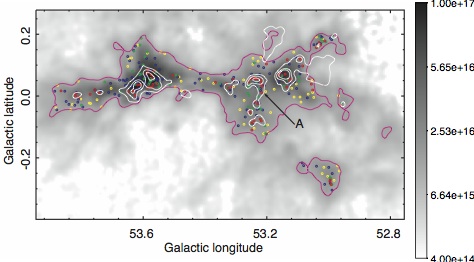}
\caption{Spatial distribution of the YSO candidates in IRDC G53.2 on 
a column density map constructed from the GRS $^{13}$CO $J=$~1--0 
image integrated at $v=$ 15--30 \kms\ (Section~2).
The map is smoothed, and 
the scale bar indicates $^{13}$CO column density of the image 
in cm$^{-2}$.
The magenta contour presents the boundary of IRDC G53.2, 
and white contours are the {\it Herschel} SPIRE 500~\um\ intensity 
contours of 3, 5, and 8~Jy~beam$^{-1}$.
Red, green, blue, and yellow symbols present 
Class I, flat-spectrum, Class II, and Class III objects, respectively.
For the arrow marked with ``A'', see Section~6.3.2.
\label{fig9}}
\end{center}
\end{figure}

As Figure~\ref{fig1} shows, bright mid-IR sources
in IRDC G53.2 are located along dark filaments in 24~\um.
Based on the classification in Section~4, 
we examine the spatial distribution of the YSO candidates in 
IRDC G53.2 in a relation to the associated CO molecular cloud 
and far-IR emission.
Figure~\ref{fig9} shows the distribution of the YSO candidates in 
each class on a $^{13}$CO column density map we construct from 
the GRS $^{13}$CO $J=$~1--0 image integrated at $v=$ 15--30~\kms\ (Section~2). 
Overall, the YSO candidates are dispersed within the boundary of 
IRDC G53.2 as drawn by the magenta contour, though they are 
also more concentrated where the $^{13}$CO column density is higher.

Column density of the molecular cloud shows a good 
correlation with far-IR continuum emission as well. 
For comparison, intensity contours from the {\it Herschel}%
\footnote{{\it Herschel} is an ESA space observatory with science instruments provided by 
European-led Principal Investigator consortia and with important participation from NASA.}
SPIRE \citep{griffin10} 500~\um\ image%
\footnote{Level 2 image retrieved from Herschel Science Archive
(http://www.cosmos.esa.int/web/herschel/science-archive)} 
with levels of 3, 5, and 8~Jy~beam$^{-1}$
are presented in white on the column density map in
Figure~\ref{fig9}, and three peak positions 
(one in the eastern and two in the western part of the IRDC) 
in both column density and 500~\um\ intensity are 
spatially coincident.
We note that there is far-IR emission outside of 
the IRDC G53.2 boundary to the north and west 
of the IRDC. This emission is probably not related to 
IRDC G53.2 but, rather, to foreground emission from CO 
clouds at another velocity observed along the same 
line of sight (see Figure~\ref{fig2}).

In Figure~\ref{fig9},
Class I objects are clustered within the 500~\um\ contours, 
particularly higher intensity levels,
whereas Class II or III objects are rather randomly distributed.
We examine the degree of clustering by comparing
the numbers of objects in each class 
within the 500~\um\ contours.
Table~\ref{tbl6} presents the numbers of the YSO 
candidates in each class within the 500~\um\ contours 
of 8, 5, 3, and 2~Jy~beam$^{-1}$.
If we compare Class I and II objects, 
$\gtrsim$40\% of the Class I objects are distributed within 5~Jy~beam$^{-1}$,
where only $\sim$22\% of the Class II objects are placed.
This indicates that 
Class I objects or YSOs in earlier classes are more 
concentrated where the 500~\um\ intensity is higher, 
although flat-spectrum objects are more widely distributed.
Clustering of Class I YSOs at high-extinction regions or the regions with 
high column density is often shown in other star-forming regions as well 
(e.g., \citealt{gutermuth09}; \citealt{myers12} and references therein;
\citealt{chavarria14}).

We also compare \ntwoone\ in each level of the 500~\um\ contours.
Flat-spectrum objects are again included with the Class I objects.
If Class I objects are concentrated in a denser region surrounded by 
more evolved YSOs, \ntwoone\ will increase as larger regions are
considered, from the densest region to the outer region, and 
there will be a gradient in the ratios. 
\citet{myers12}, using the protostar fraction, investigated age structures in well-studied 
star-forming regions, Serpens north/south clusters and the CrA cluster,
and found that there are local age variations from 0.3 to 0.9~Myr.
In Table~\ref{tbl6} we present \ntwoone\ in the 500~\um\ 
contours of each level in IRDC G53.2.
The central region with higher ($>$8~Jy~beam$^{-1}$) 500~\um\ intensity or 
higher $^{13}$CO column density shows smaller \ntwoone\ of 0.6, 
compared to 0.9 for the whole IRDC. 
The difference is small, and a local gradient among the intensity levels is 
hardly seen, likely owing to a small sample size.
Compared to the area of the IRDC, the number of YSO candidates is rather small, 
which makes a statistical comparison of the number of sources in each contour zone 
difficult.
However, we still see that Class I objects are likely concentrated along the denser filament
in IRDC G53.2.
If we apply the relation between \ntwoone\ and age we derive in
Section~6.2.3 from \citet{myers12}, 
it gives $\sim$0.2~Myr of age variation in IRDC G53.2

On the other hand,
several recent studies have shown clustering of Class I 
YSOs along filamentary structures 
or pre- and protostellar core formation along IRDCs, 
many of which are likely massive 
\citep{teixeira06,henning10,jackson10,bhavya13}.
Therefore, spatial concentration of earlier-class objects
along the bright emission in the {\it Herschel} image 
supports star formation in very early phases occurring 
in IRDC G53.2 and gives a possibility that 
a fraction of the early-phase YSOs are massive.
Detailed modeling of the objects
using a full SED including longer wavebands in the future
will be necessary to investigate physical characteristics 
of YSOs forming in IRDC G53.2.

\subsubsection{Spacings of the YSO Candidates}

In analysis of the spatial distribution of YSOs in star-forming regions,
an average spacing between sources is often used to investigate
the fragmentation processes in relation with the Jeans fragmentation 
and the subsequent dynamical evolution of the stars 
since YSOs are expected to move away from their birth sites 
as they evolve \citep{teixeira06,kumar07,winston07,gutermuth09}.
One indicator to examine spacings between YSOs is a nearest-neighbor 
distance, which is the projected distance to the nearest YSO, often
noted as NN2 distance \citep{gutermuth09}. 
In the study of the ``Spokes'' cluster in the young cluster NGC 2264, 
\citet{teixeira06} found a clear peak in their histogram of NN2 distances 
and suggested that the peak indicates the Jeans fragmentation 
of dense, dusty filaments.
A peak at small spacings with a relatively long tail of large spacings
in NN2 distance histograms is shown in young, nearby, 
star-forming clusters as well, and when the histograms show a pronounced 
peak and tail, the cumulative distributions have a steep inner slope 
and a shallow outer slope \citep[e.g., Figure 2 of][]{gutermuth09}. 

\begin{figure}
\begin{center}
\includegraphics[width=0.8\textwidth]{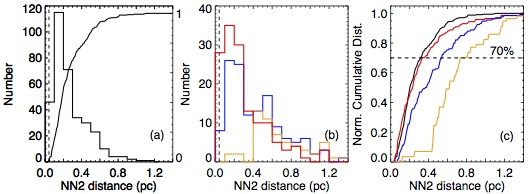}
\caption{(a) Histogram of NN2 distance of IRDC G53.2 with 
a bin size of 0.1~pc.
The dashed line marks the 5$\arcsec$ boundary below which source 
confusion becomes significant. 
The normalized cumulative distribution is overplotted.
(b) Histograms of NN2 distance for each class of YSO of IRDC G53.2.
Red, blue, and orange histograms present (Class I + flat-spectrum), Class II, 
and Class III objects, respectively. The bin size is 0.1~pc as well. 
(c) Normalized cumulative distribution of the NN2 distances of each YSO class. 
Colors are the same as those used in (b), and the black line presents all of the YSO 
candidates.
\label{fig10}}
\end{center}
\end{figure}

The histogram of the NN2 distances of the YSO candidates in 
IRDC G53.2 also shows a well-defined peak,  
as presented in Figure~\ref{fig10}(a), with a median value of 0.2~pc.  
The dashed line in the figure marks the 5$\arcsec$ (or 0.04~pc at 1.7~kpc) 
boundary below which the source confusion becomes 
significant \citep{gutermuth09}, 
which means that the most frequent spacing ($\sim$0.2 pc) is not 
an effect of resolution. 
As \citet{gutermuth09} pointed out, the cumulative distribution of the NN2 
distance has a steep slope at small spacings and a shallow slope at 
large spacings.
In Figure~\ref{fig10}(b), we compare the NN2 distances of each YSO class.
We include flat-spectrum objects in Class I, and 
they still show a clear peak at $\sim$0.2~pc. 
The NN2 distances of Class II objects also have a peak, but not as
sharp as that of Class I, 
and the NN2 distances of Class III are rather broadly distributed
at larger spacings.
We present the median of the NN2 distances of each class in 
Table~\ref{tbl7}.
The NN2 distance histograms of each class indicate that 
Class I objects are more highly clustered with smaller spacings 
(0.2~pc in median) than later classes (0.4 and 0.7~pc in median 
for Class II and Class III, respectively), and
their normalized cumulative distributions presented in Figure~\ref{fig10}(c) 
more clearly show this. 
In Figure~\ref{fig10}(c), the cumulative distribution of Class I has much 
steeper slope at small spacings than the other two classes. 
The NN2 distances that contain 70\% of the sources in each class 
are 0.3, 0.5, and 0.7~pc for Class I, II, and III, respectively 
(see Table~\ref{tbl7}). 

We perform a Kolmogorov$-$Smirnov (K-S) test to examine 
the probability that the parent distributions of each class are
the same. 
Table~\ref{tbl8} presents the K-S probabilities of the normalized 
cumulative distributions of the NN2 distances between each 
class.
The results show $< 1\%$ chance of similarity between any of 
two classes, implying that Class I objects are typically closer to 
their nearest neighbors than Class II or III objects.
Such a tendency was more obviously shown in a study on the Serpens 
cloud core, where the median NN2 distances 
of Class I, flat-spectrum, Class II, and Class III are 0.024, 0.079, 
0.097, and 0.132~pc, respectively \citep{winston07}.
An increase in the median distance between YSOs for more evolved 
sources/classes supports the idea that 
stars are born in a dense region and dispersed 
away from the birth site as they evolve,
and the smaller spacing of Class I objects in IRDC G53.2 
indicates that such a process is occurring in IRDC G53.2 as well.

\begin{figure}
\begin{center}
\includegraphics[width=0.45\textwidth]{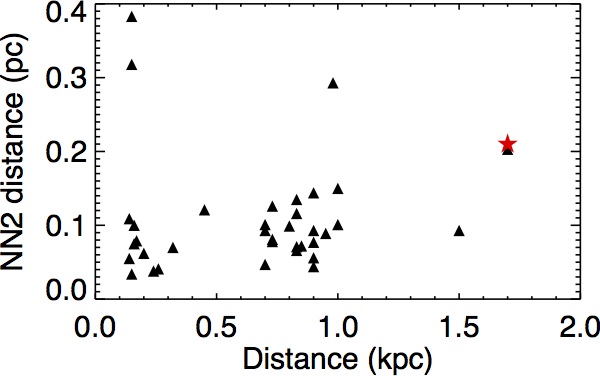}
\caption{Median NN2 distances of 36 nearby, young, star-forming clusters 
from \citet{gutermuth09} vs. their distances (black triangles). 
The median NN2 distance of IRDC G53.2 is marked with 
a red star.
\label{fig11}}
\end{center}
\end{figure}

On the other hand, 
the median NN2 distance of IRDC G53.2 is about a factor of two
larger than that of nearby star-forming clusters. 
For comparison, the mean value of the median NN2 distances of 
36 nearby, young clusters in \citet{gutermuth09} is 0.07~pc.
The relatively larger spacing of the sources in IRDC G53.2 
is likely due to its larger distance.
As discussed in Section 6.2.1, there is a selection effect 
in the sources in IRDC G53.2 from distance so that 
the NN2 distances of the sources in the IRDC can be biased to 
higher values by the smaller number of the sources compared to 
the whole area (and hence a lower surface density of sources). 
Although the mean of the median NN2 distances of the clusters 
in \citet{gutermuth09} is 0.07~pc, the median NN2 distances of the 
individual clusters are spread up to $\sim$0.4~pc, and 
they in general show a good correlation with the distance to 
the clusters as we present in Figure~\ref{fig11}.
In this plot, 
although there are a few outliers that are not easily explained,
the NN2 distance of the sources in IRDC G53.2,
lying on the tendency of the relation between 
distance and NN2 distance,
is not particularly large when one takes into account 
its distance.

In the Spokes cluster at a distance of 800~pc, 
\citet{teixeira06} found a typical separation of $\sim$0.1~pc 
for protostars distributed along its dusty filaments. 
This length scale is in very good agreement 
with the Jeans length in the cluster, so they suggested 
thermal fragmentation of the dense filamentary material.
The NN2 distance of the Spokes cluster is also smaller than 
that of IRDC G53.2. The assumption that the Spokes 
cluster is at the same distance as IRDC G53.2 does not 
change the result because \citet{teixeira06} only used 
the bright 24~\um\ sources that are detectable at 1.7 kpc.
However, the surface density of sources in the Spokes cluster
is higher owing to its smaller area than IRDC G53.2,
and we need to analyze the NN2 distance in a subregion with 
high source surface density in IRDC G53.2 to 
make a suitable comparison.
There are three subregions where the YSO candidates 
are concentrated in Figure~\ref{fig9} that are consistent with 
the 3~Jy~beam$^{-1}$ of the 500~\um\ contour, as well as dark filaments 
in 24~\um.
We derive  the NN2 distances of one of the subregions 
marked with an arrow ``A'' in Figure~\ref{fig9},
where the YSO candidates are located along the dark filament in 
24~\um\ (see Figure~\ref{fig1}).
The number of the YSO candidates in the region A is 41, 
and the median NN2 distance is $\sim$0.1~pc,
resulting in a similar length scale to that of the Spokes cluster. 

The NN2 distance in the region A is also comparable with 
the Jeans length, given as 
$\lambda_{\rm J}=(\pi c_{s}^2/G\rho_{0})^{1/2}=$ 
~0.21~pc$(T/10 {\rm K})^{1/2} \times (n_{\rm H_2}/10^4 {\rm cm}^{-3})^{-1/2}$ 
\citep{mckee07,winston07}  
for initial temperature $\sim$20 K and density $\sim10^{5} {\rm cm}^{-3}$, 
which are typical values in IRDCs \citep{pillai06,rath06,ragan11}.
Comparing with the mean core separation of $\sim$0.9~pc 
in the IRDC G011.11-0.12 \citep{henning10}, 
this indicates that IRDC G53.2, in which Jeans fragmentation is likely 
dominant, is more evolved than the IRDC G011.11-0.12.
More detailed investigation on this subregion using far-IR data and/or 
molecular line maps such as HNC (1--0) \citep[e.g.,][]{jackson10} will 
be useful to explore the fragmentation and star formation processes 
occurring in dense filamentary IRDCs.

\section{Summary and Conclusions}

We present star formation activity in IRDC G53.2, which is 
a long, filamentary IRDC at Galactic coordinates 
$(l, b) \sim (53^{\circ}.2,\, 0^{\circ}.0)$ using {\it Spitzer} 
mid-IR data. We summarize our results and give conclusions below.

1. We found that IRDC G53.2 previously identified as several 
separate IRDCs in the mid-IR images coincides precisely 
with a CO cloud at $v \sim 23$~\kms.
This gives a kinematic distance of 1.7 kpc to IRDC G53.2,
and the cloud mass determined from CO emission 
is $\sim10^5 M_{\sun}$.

2. We made a PSC of IRDC G53.2 based on
the photometry of {\it Spitzer} MIPSGAL 24~\um\ data. The finalized 
catalog after merging with the GLIMPSE Catalog contains 373 sources 
in total, including four sources saturated in the MIPS 24~\um\ 
image but listed in the {\it MSX} PSC.

3. Based on the spectral index defined in the range 2--24~\um, 
we classified the sources in the catalog.
Since IRDC G53.2 is located in the Galactic plane,
there is negligible extragalactic contamination but 
substantial field star contamination, so that 
we accounted for the field star contamination by control
field analysis.
The census of the YSO candidates in the IRDC,
if we remove the expected field star contamination, 
is determined as follows: 
77 Class I (25\%), 63 flat-spectrum (21\%), 129 Class II (42\%), 
33 Class III (11\%), and 6 No Class ($\sim$1\%) without enough 
data points to determine spectral index.

4. We compared the classification of YSO candidates
based on different classification schemes such as
spectral index, mid-IR colors, and the wavelength 
range used. Different classifications using different criteria 
generally agree well, but flat-spectrum sources show 
high uncertainty with an ambiguity between Class I and Class II. 
High extinction toward IRDC G53.2 may also
affect classification of flat-spectrum sources, but 
the effect on the overall statistics is not likely significant.

5. We compared the census of the YSO candidates in
IRDC G53.2 with those of other well-studied star-forming 
regions. A similar fraction of Class I objects in IRDC G53.2 and 
other regions indicates 
that IRDC G53.2 is similar in age or at a similar evolutionary stage 
to the nearby star-forming regions. 
On the other hand, the comparison of stellar population to 
other IRDCs shows that IRDC G53.2, with strong association 
with mid-IR stellar sources, is at a later evolutionary status among IRDCs.

6. Spatial distribution of the YSO candidates in IRDC G53.2
shows a good correlation with $^{13}$CO column density and 
far-IR emission, and
earlier-class objects tend to be more clustered
where $^{13}$CO column density is higher. 
Overall, the median distance between the YSO candidates 
and their nearest neighbors is 0.2~pc, and earlier-class 
objects have smaller spacing,
which indicates that YSOs disperse away from their birth sites 
as they evolve. 
In a small, denser region with high surface density of sources, 
the median nearest-neighbor distance 
is $\sim$0.1~pc, which is comparable to the Jeans length scale.

Characterizing star formation activity in IRDC G53.2 
based on the census of YSO candidates presented 
in this study provides an insight on the star-forming process 
occurring in IRDCs, 
particularly in an aspect of an associated molecular cloud 
rather than in an individual IRDC.
We found YSOs in various evolutionary phases, and 
our results suggest that IRDC G53.2 is an active 
star-forming region where Jeans fragmentation is 
likely dominant.
Spatial distribution of the YSO candidates, 
which has a correlation with far-IR emission 
and the objects only detected in the 24~\um\ but not 
in the IRAC bands, may imply 
the existence of much younger YSOs embedded in the IRDC. 
Further studies on full SEDs of YSO candidates 
including longer wavebands to derive mass and luminosity 
distribution of YSOs in IRDC G53.2 
will help to investigate in more detail the nature and 
star formation properties of IRDCs.

\acknowledgments
This publication makes use of molecular line data from 
the Boston University$-$FCRAO Galactic Ring Survey (GRS). 
The GRS is a joint project of Boston University and 
Five College Radio Astronomy Observatory, 
funded by the National Science Foundation under grants 
AST-9800334, AST-0098562, and AST-0100793.
This publication makes use of data products from 
the Two Micron All Sky Survey, which is a joint project of 
the University of Massachusetts and the Infrared Processing and 
Analysis Center/California Institute of Technology, funded by 
the National Aeronautics and Space Administration and 
the National Science Foundation.
This work is based on observations made with the Spitzer Space Telescope, 
which is operated by the Jet Propulsion Laboratory, California Institute of Technology, 
under a contract with NASA.
This research has made use of the SIMBAD database,
operated at CDS, Strasbourg, France.
This work was supported by NRF(National Research Foundation of Korea) 
Grant funded by the Korean Government
(NRF-2012-Fostering Core Leaders of the Future Basic Science Program).
B.-C.K. was supported by the National Research Foundation of Korea
(NRF) grant funded by the Korea Government (MSIP) (No. 2012R1A4A1028713).



\clearpage

\clearpage

\clearpage

\begin{deluxetable}{lr}
\tabletypesize{\small}
\tablecaption{Parameters of IRDC G53.2 \label{tbl1}}
\tablewidth{0pt}
\tablehead{
\colhead{Parameter} & {IRDC G53.2 }
}
\startdata
Center $(l,b)$ & $\sim (53^{\circ}.21, -0^{\circ}.086)$ \\
Angular Size  & $\sim 1^{\circ}.5 \times 0^{\circ}.7 $\\
Linear Size & $45~{\rm pc} \times 21~{\rm pc}$ \\
$v_{/rm center}$  & $+22.9 \pm 1.0$ (\kms)  \\
$\Delta v_{rm width}$  & $1.7 \pm 0.5$ (\kms) \\
Kinematic distance  & 1.7 (kpc) \\
Mass & $6.2\times10^4$\tablenotemark{a}, 
             $1\times10^5$\tablenotemark{b} (\msol) \\
Surface density  & 0.076\tablenotemark{a},
                              0.033\tablenotemark{b} (g cm$^{-2}$)\\
\enddata
\tablenotetext{a}{The boundary of IRDC G53.2 for the analysis 
of the star formation activity in mid-IR (the $^{13}$CO integrated 
intensity contour at $\int T_{mb} dv=6.3$ K \kms).}
\tablenotetext{b}{The cloud boundary accounting for its low-density envelope 
(at $\int T_{mb} dv=2.1$ K \kms). See Section~2.}
\end{deluxetable}
\clearpage
\begin{deluxetable}{ccccccccccccr}
\rotate
\tabletypesize{\tiny}
\setlength{\tabcolsep}{0.07in}
\tablecaption{
Point-source Catalog of IRDC G53.2 \label{tbl2}}
\tablewidth{0pt}
\tablehead{
\colhead{No} & \colhead{R.A. (J2000)} & \colhead{Decl. (J2000)} & 
\colhead{\it J} & \colhead{\it H} & \colhead{\it K$_s$} &
\colhead{[3.6]} & \colhead{[4.5]} & \colhead{[5.8]} & \colhead{[8.0]} &
\colhead{[24]} & \colhead{$\alpha$}\tablenotemark{a} & 
\colhead{Class}\tablenotemark{b}
}
\startdata
1 & 
19:30:19.59 & 18:18:17.21 & 
$11.84\pm0.022$ & $11.11\pm0.022$ & $10.72\pm0.021$ & 
$10.57\pm0.156$ & $10.26\pm0.181$ & \nodata & 
\nodata & $3.27\pm0.044$ & 
\nodata & I$^{d}$ \\
2 & 
19:30:22.95 & 18:20:27.92 & 
\nodata & \nodata & $13.83\pm0.076$ & 
$10.01\pm0.214$ & $7.39\pm0.159$ & $5.69\pm0.026$ & 
$4.94\pm0.025$ & $0.82\pm0.061$ & 
$1.77\pm0.57$ & I \\
3 & 
19:30:22.14 & 18:20:46.43 & 
$14.83\pm0.049$ & $12.71\pm0.034$ & $11.46\pm0.031$ & 
$9.67\pm0.061$ & $8.84\pm0.063$ & $8.23\pm0.060$ & 
$7.80\pm0.183$ & $5.62\pm0.046$ & 
$-0.74\pm0.19$ & II \\
4 & 
19:30:14.05 & 18:18:22.50 & 
$12.91\pm0.025$ & $11.33\pm0.022$ & $10.27\pm0.018$ & 
$9.09\pm0.039$ & $8.65\pm0.047$ & $8.15\pm0.036$ & 
$7.30\pm0.029$ & $4.16\pm0.045$ & 
$-0.55\pm0.05$ & II$^{a}$ \\
5 & 
19:30:17.10 & 18:19:43.00 & 
$11.97\pm0.025$ & \nodata & \nodata & 
$6.44\pm0.063$ & \nodata & $5.64\pm0.036$ & 
$5.21\pm0.025$ & $3.44\pm0.045$ & 
$-1.30\pm0.08$ & II$^{a}$ \\
6 & 
19:30:25.17 & 18:23:24.18 & 
$11.61\pm0.020$ & $8.91\pm0.027$ & $7.43\pm0.022$ & 
$5.74\pm0.062$ & \nodata & $5.09\pm0.026$ & 
$4.84\pm0.027$ & $3.79\pm0.044$ & 
$-1.58\pm0.23$ & II$^{a}$ \\
7 & 
19:30:12.58 & 18:18:13.46 & 
\nodata & $13.95\pm0.040$ & $10.54\pm0.020$ & 
$6.76\pm0.032$ & $6.20\pm0.103$ & $4.76\pm0.027$ & 
$4.22\pm0.026$ & $2.73\pm0.044$ & 
$-0.19\pm0.53$ & Flat$^{a}$ \\
8 & 
19:30:20.61 & 18:21:34.20 & 
\nodata & $14.33\pm0.057$ & $13.09\pm0.043$ & 
$11.18\pm0.090$ & $10.68\pm0.077$ & $10.01\pm0.115$ & 
\nodata & $6.30\pm0.044$ & 
$-0.37\pm0.12$ & II \\
\enddata
\tablenotetext{a}{Defined as $\alpha = d~{\rm log}\,(\lambda\,F_{\lambda})/d~{\rm log}\,
(\lambda)$ \citep{lada87} between 2 and 24~\um.}
\tablenotetext{b}{a = IRAC magnitudes are from the GLIMPSEI Archive;
d = a deeply embedded protostar}
\end{deluxetable}

\clearpage

\begin{deluxetable}{ccccccccccccr}
\rotate
\tabletypesize{\tiny}
\setlength{\tabcolsep}{0.07in}
\tablecaption{
Saturated Sources in 24~\um \label{tbl3}}
\tablewidth{0pt}
\tablehead{
\colhead{No} & \colhead{R.A. (J2000)} & \colhead{Decl. (J2000)} & 
\colhead{\it J} & \colhead{\it H} & \colhead{\it K$_s$} &
\colhead{[3.6]} & \colhead{[4.5]} & \colhead{[5.8]} & \colhead{[8.0]/{\it MSX} A} &
\colhead{{\it MSX} E} & \colhead{$\alpha$} & \colhead{Class}
}
\startdata
1 & 
19:29:17.59 & 17:56:22.81 & 
\nodata & \nodata & $11.92\pm0.073$ & 
$7.03\pm0.106$ & $4.42\pm0.110$ & $3.54\pm0.050$ & 
$2.36\pm0.045$ & $-1.39\pm0.065$ & 
$1.88\pm0.62$ & I \\
2 & 
19:29:17.29 & 17:56:17.59 & 
\nodata & $14.85\pm0.181$ & $11.97\pm0.091$ & 
$8.88\pm0.204$ & $7.49\pm0.162$ & $6.19\pm0.046$ & 
$4.97\pm0.044$ & ... & 
$2.21\pm0.13$ & I \\
3 & 
19:29:41.28 & 18:06:39.38 & 
$8.19\pm0.027$ & $5.83\pm0.023$ & $4.63\pm0.020$ & 
\nodata & $2.85\pm0.165$ & $2.55\pm0.035$ & 
$1.78\pm0.045$ & $0.17\pm0.066$ & 
$-1.11\pm0.09$ & II \\
4 & 
19:30:01.68 & 18:19:14.23 & 
$8.73\pm0.032$ & $6.23\pm0.020$ & $4.81\pm0.021$ & 
$4.45\pm0.160$ & \nodata & \nodata & 
$1.46\pm0.045$ & $-0.51\pm0.065$ & 
$-0.59\pm0.20$ & II \\
\enddata
\tablecomments{Only No. 2 is detected in GLIMPSE catalog/archive at 8 \um.
Others use {\it MSX} A magnitudes.}
\end{deluxetable}

\clearpage

\begin{deluxetable}{lcccc}
\tabletypesize{\small}
\tablewidth{0pt}
\tablecaption{
Comparison of Classified Sources in IRDC G53.2 and Control Fields 
\label{tbl4}}
\tablehead{
\colhead{Class} & \colhead{Number in} & 
\colhead{Number Scaled to} & \colhead{Number Identified in} &
\colhead{Contamination} \\
\colhead{ } & \colhead{Control Fields} &
\colhead{Area of IRDC G53.2} & 
\colhead{IRDC G53.2} & \colhead{Fraction}
}
\startdata
 I	& 2 & 0.9 & 76(2) & 1.2\% \\
Flat-spectrum	& 6 & 2.8 & 66 & 4.2\% \\
 II	& 13 & 6.1 & 133(2) & 4.6\% \\
III	& 61 & 28.5 & 62 & 45.9\% \\
Photospheric & 136 & 63.5 & 26 & $\sim$100\% \\
\enddata
\tablecomments{The sources of IRDC G53.2 (column 4) only account for
the sources in Table~\ref{tbl2}, and the numbers of the MIPS-saturated 
sources (Table~\ref{tbl3}) are presented in parentheses for reference. 
Note that including the MIPS-saturated sources does not 
affect the fraction of field star contamination (column 5).}
\end{deluxetable}

\begin{deluxetable}{lcccc}
\tabletypesize{\small}
\tablewidth{0pt}
\tablecaption{
Comparison of Number Ratio of Class II to Class I Objects (\ntwoone) 
between IRDC G53.2 and Other Star-forming Regions 
\label{tbl5}}
\tablehead{
\colhead{Region} & \colhead{Distance (kpc)} & \colhead{Without Correction} & 
\colhead{\aall-based Class} & 
\colhead{At 1.7 kpc\tablenotemark{a}}
}
\startdata
IRDC G53.2\tablenotemark{b}  & 1.7  & 0.9 & 0.9 & 0.9 \\
\citet{gutermuth09} & $<$1 & 3.7 & 2.0 & 0.8 \\
c2d clouds\tablenotemark{c} & $<$0.3 & 2.1 & 2.1 & 1.1 \\
Vul OB1\tablenotemark{d} & 2.3 & 1.9 & 1.2 & $\sim$1.2\tablenotemark{e}
\enddata
\tablenotetext{a}{Classification is based on \aall.}
\tablenotetext{b}{The numbers are after removing field star contamination.}
\tablenotetext{c}{\cite{evans09}}
\tablenotetext{d}{\cite{billot10}}
\tablenotetext{e}{Since Vul OB1 is at the similar distance to IRDC G53.2,
we use \ntwoone\ without correction.}
\end{deluxetable}

\begin{deluxetable}{cccccc}
\tabletypesize{\small}
\tablewidth{0pt}
\tablecaption{Numbers of Young Stellar Object Candidates 
in IRDC G53.2 in the {\it Herschel} 500~\um\ Contours
\label{tbl6}}
\tablehead{
\colhead{Class} & \colhead{8 Jy~beam$^{-1}$} & 
\colhead{5 Jy~beam$^{-1}$ } & \colhead{3 Jy~beam$^{-1}$} &
\colhead{2 Jy~beam$^{-1}$} & \colhead{all IRDC\tablenotemark{a}}
}
\startdata
 I	  & 15 (19\%) & 33(42\%) & 56(72\%) & 74(95\%) & 78(100\%) \\
Flat-spectrum  & 4 (6\%) & 18 (27\%) & 36 (55\%) & 53 (80\%) & 66 (100\%) \\
 II  & 11 (8\%) & 30 (22\%) & 54 (40\%) & 84 (62\%) & 135 (100\%) \\
 III & 0 (0\%) & 0 (0\%) & 2 (3\%) & 15 (24\%) & 62 (100\%) \\
\tableline
\ntwoone\tablenotemark{b} & 0.6 & 0.6 & 0.6 & 0.7 & 0.9
\enddata
\tablenotetext{a}{The numbers without removing field star contamination.}
\tablenotetext{b}{Flat-spectrum is included in Class I.}
\end{deluxetable}

\begin{deluxetable}{lccc}
\tablecaption{NN2 Distances of YSO Candidates in IRDC G53.2 
\label{tbl7}}
\tablewidth{0pt}
\tablehead{
\colhead{Class} & \colhead{I+Flat} & 
\colhead{II} & \colhead{III}
}
\startdata
Median NN2 distance (pc) & 0.24 & 0.38 & 0.68 \\
70\% of cumulative dist. (pc)\tablenotemark{a} & 0.33 & 0.54 & 0.73 \\
80\% of cumulative dist. (pc)\tablenotemark{a} & 0.46 & 0.68 & 0.98
\enddata
\tablenotetext{a}{NN2 distance where 70\% or 80\% of the sources 
are included in normalized cumulative distribution.}
\end{deluxetable}

\begin{deluxetable}{cccc}
\tablecaption{Kolmogorov$-$Smirnov Probabilities of the Normalized 
Cumulative Distributions of the NN2 Distances between Each 
YSO Class
\label{tbl8}}
\tablewidth{0pt}
\tablehead{
\colhead{Class} & \colhead{I+Flat} & 
\colhead{II} & \colhead{III}
}
\startdata
I+flat & $\cdots$ & $9\times10^{-3}$ & $8\times10^{-12}$ \\
II &  $9\times10^{-3}$  & $\cdots$ &  $1.7\times10^{-4}$ \\
III &  $8\times10^{-12}$ & $1.7\times10^{-4}$ & $\cdots$
\enddata
\end{deluxetable}


\end{document}